\documentclass[prl,twocolumn, amsmath,amssymb,superscriptaddress,nofootinbib,11pt]{revtex4-2}
\usepackage{url}

\usepackage{epsfig}
\usepackage{amsfonts}
\usepackage{graphicx}
\usepackage{subfigure}
\usepackage{epsfig}
\usepackage{eepic}
\usepackage{amsmath}
\usepackage{amssymb}
\usepackage{color}
\usepackage{bbm}
\usepackage{dcolumn}
\usepackage{bm}
\usepackage[normalem]{ulem}
\usepackage{mathrsfs}
\usepackage{bbold}
\usepackage{datetime}

\usepackage{overpic} 
\usepackage{rotating}
\usepackage[usenames,dvipsnames]{xcolor}
\usepackage[colorlinks=true,citecolor=Magenta,linkcolor=Green,urlcolor=Green]{hyperref}
\usepackage{lipsum} 
\usepackage{tikz,tikz-3dplot}
\usetikzlibrary{shapes.geometric}
\usepackage{etoolbox} 
\usepackage[capitalize]{cleveref}
\usepackage{extarrows}
\usepackage{ragged2e}

\def\bc{\begin{center}}
\def\nno{\nonumber}
\def\ec{\end{center}}
\def\be{\begin{eqnarray}}
\def\ee{\end{eqnarray}}
\definecolor{dyellow}{rgb}{1.,0.8,.0}
\definecolor{myblue}{rgb}{.1,.1,.7}
\definecolor{dcyan}{rgb}{.0,.6,.6}
\definecolor{dmagenta}{rgb}{0.6,0.0,0.6}
\definecolor{brown}{rgb}{0.6,0.2,0.}
\definecolor{darkblue}{rgb}{.0,.0,0.5}
\definecolor{darkred}{rgb}{0.75,0.0,0.0}
\definecolor{orange}{rgb}{1.,.6,.0}
\definecolor{dorange}{rgb}{0.8,.4,.0}
\definecolor{darkgreen}{rgb}{0.0,0.6,0.0}
\definecolor{purple}{rgb}{.4,.0,.4}
\definecolor{lightgrey}{rgb}{0.7, 0.7, 0.7}
\definecolor{grey}{rgb}{0.4, 0.4, 0.4}

\usepackage{geometry}
\usepackage[font=small,labelfont=bf,justification=raggedright]{caption}

\newcommand{\xdownarrow}[1]{%
  {\left\downarrow\vbox to #1{}\right.\kern-\nulldelimiterspace}
}
\newcommand{\xuparrow}[1]{%
  {\left\uparrow\vbox to #1{}\right.\kern-\nulldelimiterspace}
}

\begin{document}
\title{Asymmetric Symmetry Breaking:\\ Unequal Probabilities of Vacuum Selection}
\author{Tian-Chi Ma}\email{tianchima@buaa.edu.cn}
\affiliation{Center for Gravitational Physics, Department of Space Science, Beihang University, Beijing 100191, China}
\author{Han-Qing Shi}\email{by2030104@buaa.edu.cn}\thanks{Corresponding author}
\affiliation{Center for Gravitational Physics, Department of Space Science, Beihang University, Beijing 100191, China}
\author{Hai-Qing Zhang\href{https://orcid.org/0000-0003-4941-7432}{\includegraphics[scale=0.05]{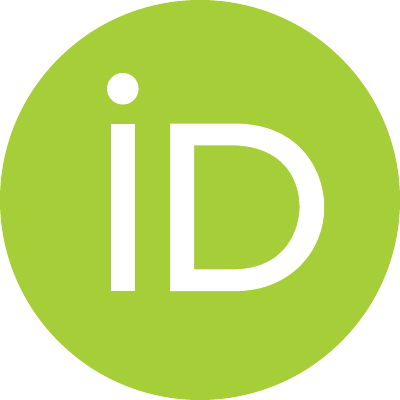}}}\email{hqzhang@buaa.edu.cn}\thanks{Corresponding author}
\affiliation{Center for Gravitational Physics, Department of Space Science, Beihang University, Beijing 100191, China}
\affiliation{Peng Huanwu Collaborative Center for Research and Education, Beihang University, Beijing 100191, China}

\begin{abstract}
Spontaneous symmetry breaking is a fundamental notion in modern physics, ranging from high energy to condensed matter. However, the usual spontaneous symmetry breaking only considers the equal probability to select the vacua. In this work, we conceive a model to realize the unequal probability of the symmetry breaking, leading to an unbalanced number of ground states. Specifically, we study the probabilities of a scalar field to roll down from the top of a potential, where the top is only $C^1$ continuous. As the whole system is subject to random perturbations, we find that the probability for the field to roll down to the left or right side depends the square root of the second derivative of the potential at the top. We solve this problem theoretically by using the Fokker-Planck equations in stochastic process and verify our findings numerically. This study may potentially be a new mechanism to explain the origins of asymmetries in the Universe. 
\end{abstract}

\maketitle


Spontaneous symmetry breaking (SSB) is an important concept in contemporary physics. It is vital for the Higgs mechanism in particle physics \cite{Higgs:1964ia,Higgs:1964pj}, the cosmological phase transitions in cosmology \cite{Guth:1980zm}, the superconducting phenomena in condensed matter physics \cite{tinkham2004introduction} and the formation of topological defects in nonequilibrium dynamics \cite{Kibble76a,Kibble76b,Zurek96a}, etc. A typical picture of SSB is a field to roll down a potential with Mexican-hat profile. The field was sitting at the top of the potential initially with higher symmetries. Since the top is unstable, the field will then roll down the potential randomly and settle down in one of the lowest energy states, aka ground states. Compared to the symmetries at the top, the chosen state has lower symmetries, thus, symmetry was broken. We call this process of SSB as ``vacuum selection''. 

Commonly, the probabilities for the field to roll down the potential are identical, that is to say, the probabilities of the vacuum selection are equal. For instance, if the potential is only one-dimensional, then the filed to roll down from the top to the left or right side are exactly one-half. A natural question is: how can it have unequal probabilities for the vacuum selection? This can be equivalently expressed as how can the field have the unequal probabilities to roll down the potential.  If it can be realized, it may have important meanings to the physics of symmetry breaking, since ground states are overarching in quantum field theory \cite{Weinberg:1995mt,Weinberg:1996kr}. 

\begin{figure}[h]
\centering
\includegraphics[trim=0cm 0cm 0cm 0cm, clip=true, scale=0.2, angle=-90]{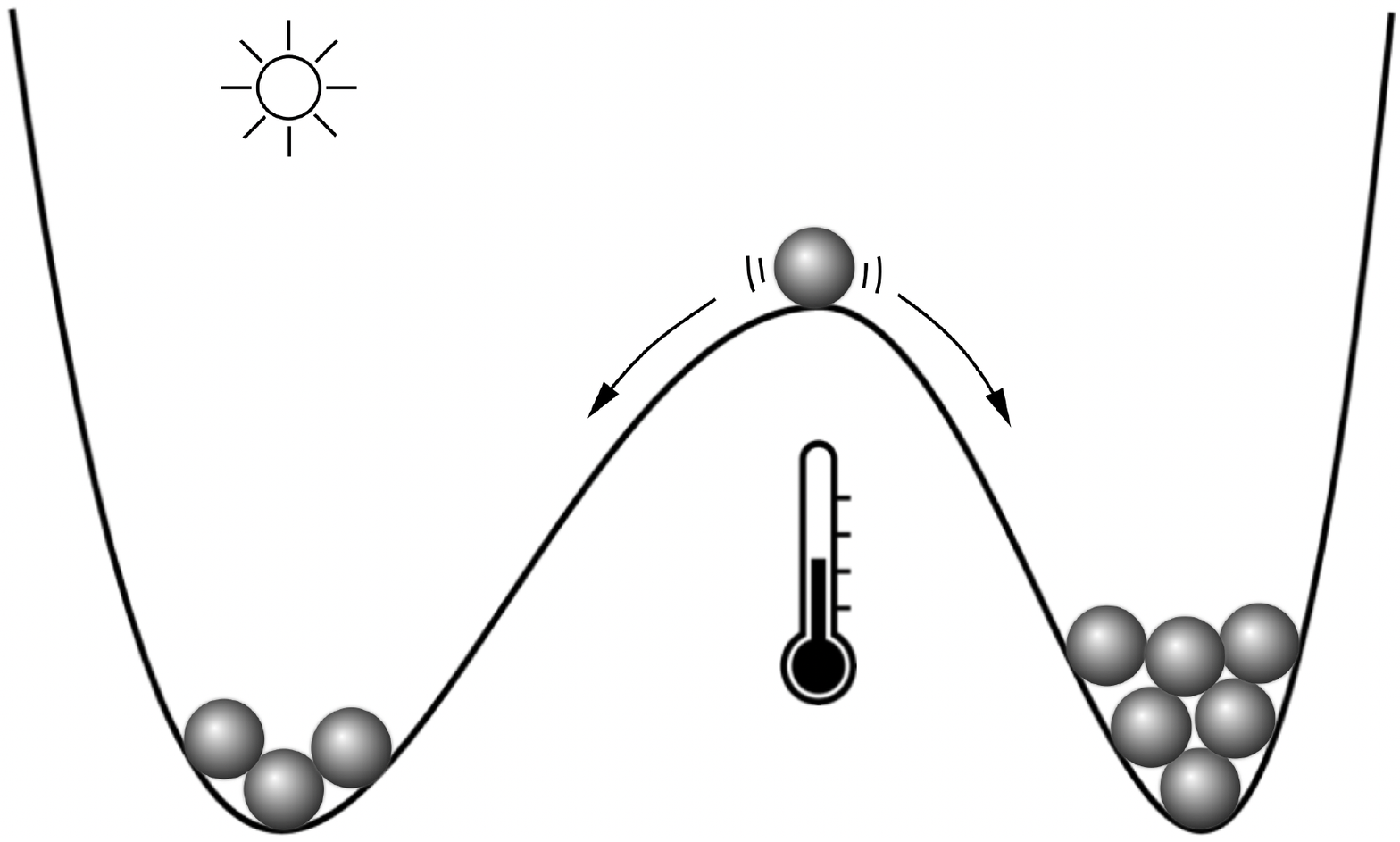}
\caption{Schematic picture to illustrate the rolling down of the fields. The system is subject to thermal perturbations, and the top of the potential is only $C^1$ continuous. Fields tend to roll down to the steeper side. }\label{potential}
\end{figure}

In this work, we utilize an artificial Higgs-type potential to realize the unequal probabilities of the vacuum selection. For simplicity, we will only consider one-dimensional potential. In particular, the top of the potential is only $C^1$ continuous, but discontinuous at higher order, like the sketchy picture shows in Fig.\ref{potential}. We call this potential as ``$\mathring{C}^1$ continuous'', in which the upper circle indicates that only the top is $C^1$ continuous while other parts are $C^\infty$ continuous. 

We assume that the whole system is subject to random external forces, for example coupled to a thermal reservoir. Under the random perturbations, the field at the top will have the probability to roll down to the left or right side. And eventually settle down to the lower energy states at the minimum of the potential. In this work, we use the Fokker-Planck equation in stochastic process \cite{Risken1996} to solve this problem theoretically, and find that the probability for the field to roll down to the left (right) depends on the square root of the second derivative of the potential at the left (right) side of the top. Then we adopt the time-dependent Ginzburg-Landau model \cite{hohenberg2015introduction} for real scalar fields to simulate this process by virtue of the Higgs-type potential. From this model, we see that the numerical results are consistent with the theoretical predictions very well. 


\section{Theory}

 We consider the time-dependent Ginzburg-Landau equation with an external random force $F(t)$,
\be\label{GL}
\frac{d\phi}{dt}+\Gamma\frac{\partial V(\phi)}{\partial \phi}=F(t)
\ee
where $\phi$ is a real scalar field, $\Gamma$ is a coefficient related to the dissipations and $V(\phi)$ is the potential. The random force $F(t)$ satisfies the Gaussian white noise relation $\langle F(t)\rangle=0, \langle F(t)F(t')\rangle=Q\delta(t-t')$ where $Q$ is the fluctuation strength. If the random perturbation comes from the thermal fluctuations of a heat bath, $Q$ will be linearly proportional to the temperature $T$ due to the fluctuation-dissipation theorem \cite{Kubo:1966fyg}, i.e., $Q=2\Gamma T$ where we have set the Boltzmann constant $k_B\equiv1$. 

From stochastic process, the probability $P(\phi, t)$ to find the field in the range $[\phi, \phi+d\phi]$ at some instant $t$ satisfies the Fokker-Planck  equation \cite{Risken1996},
\be\label{FPGL}
\frac{\partial P}{\partial t}=\Gamma\frac{\partial}{\partial \phi}\left(\frac{\partial V}{\partial\phi}P\right)+\frac{Q}{2}\frac{\partial^2 P}{\partial\phi^2}.
\ee
Assume that at initial time the field is sitting at the top of the potential like in Fig.\ref{potential}, because of the random fluctuations the field is unstable and will eventually roll down the potential to the left or right side. Without loss of generality, we can always translate the top of the potential to be seated at $\phi=0$. Intuitively, the probability for the field to roll down to the left (right) of potential will only depends on the profile of the top and its vicinity, and has nothing to do with the far-away regions from the top.  Therefore, we can expand the potential near the top $\phi=0$, i.e., $V(\phi)\approx V_0+V'_0\phi+\frac12V''_0\phi^2+\dots$, where the symbol $^\prime$ indicates the derivative of $V$ with respect to $\phi$ and the subscripts $0$ imply that the values are taken at $\phi=0$. Since the top is a maximum, we get $V'_0=0$. Hence, $\partial V/\partial\phi\approx V_0''\phi+\cdots$. Therefore, the corresponding Fokker-Planck equation Eq.\eqref{FPGL} reduces to
\be\label{expandFP}
\frac{\partial P}{\partial t}=\Gamma V''_0\frac{\partial}{\partial\phi}\left(\phi P\right)+\frac{Q}{2}\frac{\partial^2 P}{\partial\phi^2}
\ee
In the Eq.\eqref{expandFP} we have ignored the higher terms of the expansion in $\phi$ since they make minor contributions compared to the linear term of $\phi$ near $\phi=0$. 

Imagine that we make plenty of experiments to roll down the fields from the top, after enough times of the experiments we can find that the probability of rolling down to the left (right) of the potential will reach a stationary value, i.e, $\partial P/\partial t=0$.  From the reduced Fokker-Plank equation Eq.\eqref{expandFP}, the stationary solution is 
\be\label{probFP}
P_{\rm st.}(\phi)=\sqrt{\frac{\Gamma |V''_0|}{\pi Q}}e^{-\frac{\Gamma V''_0 \phi^2}{Q}}. 
\ee
It should be noted that the above solution is only valid nearby the top of the potential. 

 At the initial time $\phi$ sits at the top of the potential where $\phi=0$. Then we evolve the system to see whether $\phi$ will roll down to the left or right. Random fluctuations of $F(t)$ in this case plays a role of disturbing $\phi$ from $\phi=0$ to $\phi=\epsilon$, where $\epsilon$ is a relatively small value. If $\epsilon$ is negative (positive), the field $\phi$ will roll to the left (right) and then roll down the potential persistently until the left (right) minimum. \footnote{This behavior can be intuitively seen in the inset plot of the following Fig.\ref{pzpf} {\bf b}.} Since $\epsilon$ is small, we can ignore the exponential term in the Eq.\eqref{probFP}. Therefore, we can speculate that the probability to roll down to the left (right) directly depends the factor $\sqrt{{\Gamma |V''_0|}/{(\pi Q)}}$. Now consider that the potential is only $\mathring{C}^1$ continuous, then the second derivative of the potential $V''_0$ near $\phi\sim0$ are different on each side of the top. We can define $V''_\pm=\partial^2 V/\partial\phi^2$ at $\phi=0_\pm$, and then get
\be\label{pzoverpf}
\frac{P_+}{P_-}=\frac{\sqrt{{\Gamma |V''_+|}/{(\pi Q)}}}{\sqrt{{\Gamma |V''_-|}/{(\pi Q)}}}=\sqrt{\frac{V''_+}{V''_-}}.
\ee
where $P_- (P_+)$ indicates the probability of rolling down to the left (right) side. We have omitted the symbols of absolute value since the ratio $V''_+/V''_-$ are positive although they are individually negative. Since $P_++P_-=1$, it is readily to obtain from the Eq.\eqref{pzoverpf} that 
\be\label{pzandpf}
P_+=\frac{\sqrt{V''_+/V''_-}}{1+\sqrt{V''_+/V''_-}}. 
\ee
Or, equivalently, $ P_-=\frac{\sqrt{V''_-/V''_+}}{1+\sqrt{V''_-/V''_+}}$. 

Therefore, we see that the probability to roll down left or right depends on the square root of the ratio between the second derivatives of the potential at the top. Eqs.\eqref{pzoverpf} and \eqref{pzandpf} are our main theoretical findings to predict the probabilities for the vacuum selection in the $\mathring{C}^1$ continuous potential.

\section{Numerical Evidences}
In the following we will use an effective potential to verify our conclusions, i.e., Eq.\eqref{pzoverpf} and \eqref{pzandpf} above. Specifically, the potential is the Higgs-type potential as, 
\be \label{higgs}
V(\phi)&=&\theta(-\phi)(-\alpha_-\phi^2+\phi^4)\nno\\
&&+\theta(\phi)(-\alpha_+\phi^2+\phi^4), 
\ee
where $\theta$ is the Heaviside step function with 
\be
\theta(x)=\left\{\begin{array}{cc}
 1,& x>0\\
\frac12,&x=0\\
0,&x<0.
\end{array}\right.
\ee
One may find that the Higgs-type potential is not symmetric under $\phi\to-\phi$, therefore the original system is not symmetric under $Z_2$ transformation. Strictly speaking, this is not even a symmetry breaking. However, we need to stress that our starting point is from the time-dependent Ginzburg-Landau equation rather than from the Lagrangian or Hamiltonian formalism. We adopt the effective asymmetric potential to learn the probabilities of the field to roll down to the left or right. Nevertheless, in order to compare our study to the ordinary ``symmetry breaking'' with equal probabilities, we deliberately call our case as ``asymmetric symmetry breaking''. The meaning of it can be understood from this aspect. 

\begin{figure}[t]
\centering
\includegraphics[trim=4.6cm 9.4cm 4.5cm 9.8cm, clip=true, scale=0.39, angle=0]{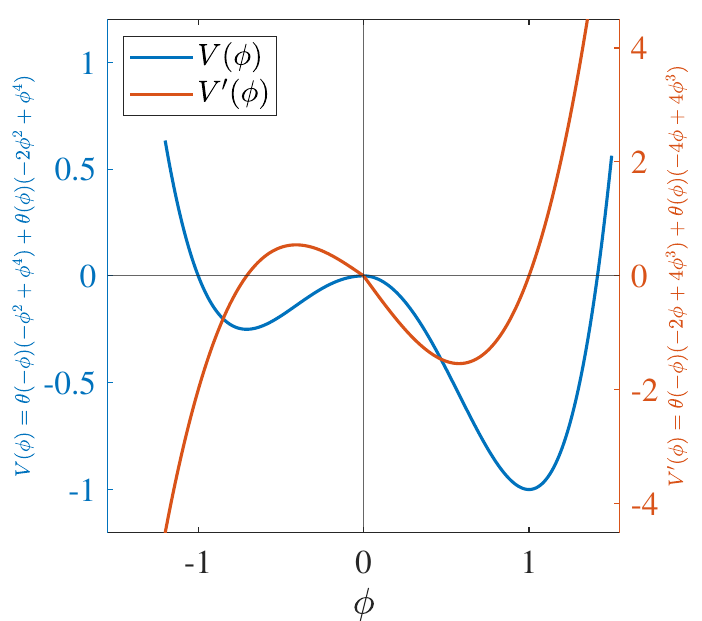}
\put(-128,110){\bf a}
\includegraphics[trim=4.6cm 9.4cm 5.5cm 9.8cm, clip=true, scale=0.39, angle=0]{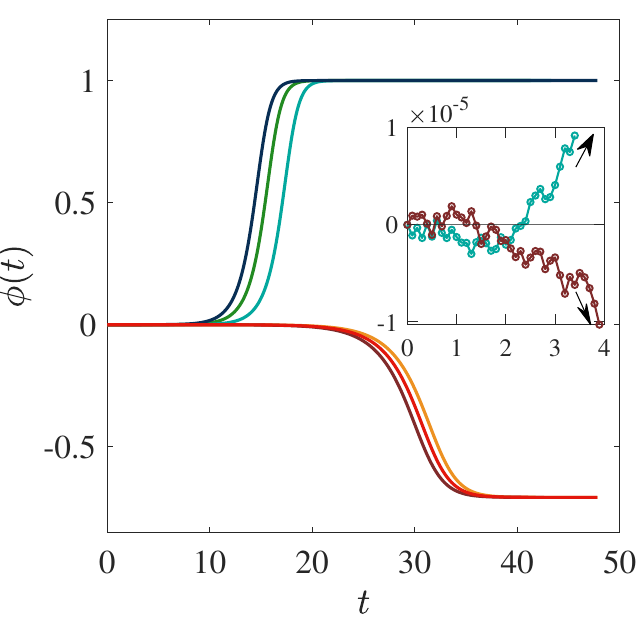}
\put(-118,110){\bf b}\\
\includegraphics[trim=4.7cm 9.4cm 5.5cm 9.8cm, clip=true, scale=0.39, angle=0]{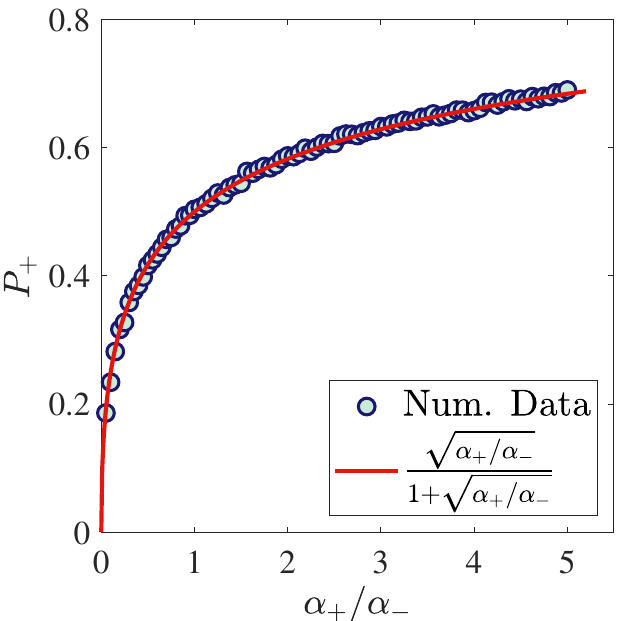}
\put(-118,110){\bf c}~~~~~
\includegraphics[trim=4.7cm 9.4cm 5.5cm 9.8cm, clip=true, scale=0.39, angle=0]{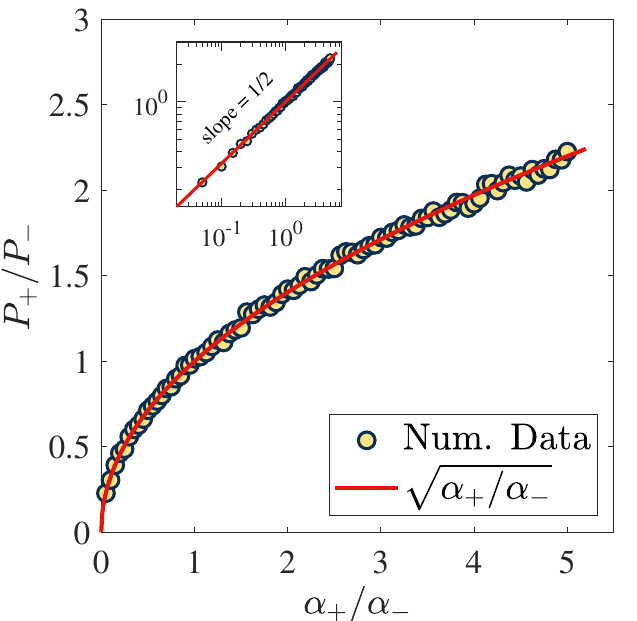}
\put(-118,110){\bf d}
\caption{{\bf a}. The Higgs-type potential $V(\phi)$ in Eq.\eqref{higgs} with $\alpha_-=1$ and $\alpha_+=2$, and its derivative $V'(\phi)$; {\bf b}. Time evolutions of the scalar fields indicating the rolling down from the top to the left and right minimum.  The inset plot shows the very early fluctuations around the top due to the temperature fluctuations. After a while, they will go away from the top and persistently roll down the potential as the arrows indicate; {\bf c}. Relation between the probability $P_+$ to the ratio $\alpha_+/\alpha_-$. Numerical results match the theoretical prediction in Eq.\eqref{pzandpf} very well; {\bf d}. Relations between the ratio $P_+/P_-$ to the ratio $\alpha_+/\alpha_-$. Numerical results are completely consistent with the theoretical prediction in Eq.\eqref{pzoverpf}. The inset plot shows the corresponding double logarithmic plot, in which the slope $1/2$ exactly indicates the square root relations in Eq.\eqref{pzoverpf}. }\label{pzpf}
\end{figure}

In numerics, we fix $\alpha_-\equiv1$ and vary $\alpha_+$ in the range of $\alpha_+\in[0.05, 5]$. Please find the profile of the potential (blue) and its derivative (red) in the Fig.\ref{pzpf} {\bf a}, in which we set $\alpha_+=2$. Obviously, the profile of the potential satisfies the requirement of $\mathring{C}^1$ continuous. This potential has two minima $\phi_+=1$ and $\phi_-=-1/\sqrt2$ which are exactly the positive and negative vacuum values. In the numerics, we use the 4-th Runge-Kutta methods to evolve the system. And we have set the time step as $\Delta t=0.1$, the coefficient $\Gamma=5$ and the temperature $T=0.01$. For the counting statistics, we independently simulate the dynamics for 30000 times. 

In the panel {\bf b} of Fig.\ref{pzpf} we show the time evolution of the scalar fields with the potential in panel {\bf a}. In this plot we show three cases for the scalar field to select the positive vacuum and three cases for selecting the negative vacuum. Because of the random temperature perturbations, each three lines are not completely overlap. But at the final time they will overlap and reach the equilibrium values of the positive (negative) vacuum, respectively. The positive (negative) equilibrium value is $1$ ($-0.707$) which are exactly the value of $\phi_+=1$ ($\phi_-=-1/\sqrt2$) corresponding to the two minima of the potential. In the inset plot of panel {\bf b} we show the early evolutions of two cases of the fields. We can see that in the very early time the fields will fluctuate around $\phi=0$ due to the external temperature perturbations. But this fluctuation near the top of the potential will not last long enough time because they are unstable at the top. At certain time they will go away from the top at $\phi=0$, and then persistently roll down the hill until meeting the minimum of the potential. The arrows in the inset plot imply the persistence of rolling down the hill. Finally, they will settle down at the minima of the potential, which is reflected by staying at the plateaus in panel {\bf b}.  

Panel {\bf c} of Fig.\ref{pzpf} shows the relation between $P_+$ and $\alpha_+/\alpha_-$. From the potential Eq.\eqref{higgs} we see that $\alpha_+/\alpha_-={V''_+/V''_-}$ at the top of the potential. From panel {\bf c} we see that the numerical results (circles) match the theoretical prediction (red line) $P_+=\sqrt{V''_+/V''_-}/\left(1+\sqrt{V''_+/V''_-}\right)$ very well. Fig.\ref{pzpf} {\bf c} demonstrates the theoretical prediction in Eq.\eqref{pzandpf}. 

Panel {\bf d} of Fig.\ref{pzpf} exhibits the relation between $P_+/P_-$ and $\alpha_+/\alpha_-$. Therefore, we see that the numerical data (circles) also match the theoretical prediction (red line) $P_+/P_-=\sqrt{V''_+/V''_-}$ very well. The inset plot in panel {\bf d} shows the double logarithmic plots of the relation, and the slope $1/2$ exactly means the square root relation in $P_+/P_-=\sqrt{V''_+/V''_-}$. Fig.\ref{pzpf} {\bf d} verifies the theoretical prediction in Eq.\eqref{pzoverpf}.

\subsection{Counting statistics of the binomial distribution}
If we regard the rolling down to the right side as a success with probability $P_+$, then rolling down to the left is a failure with probability $P_-=1-P_+$. Therefore, each time of the rolling down is a Bernoulli trial. Then after $n$ times of trials, the probability of rolling down to the right side as $k$ times should satisfy the binomial distributions \cite{2002Binomial}
\be\label{binodist}
B(n,k)=\binom{n}{k}P_+^k(1-P_+)^{n-k}.
\ee

\begin{figure}[t]
\centering
\includegraphics[trim=4.6cm 9.4cm 5.9cm 9.cm, clip=true, scale=0.39, angle=0]{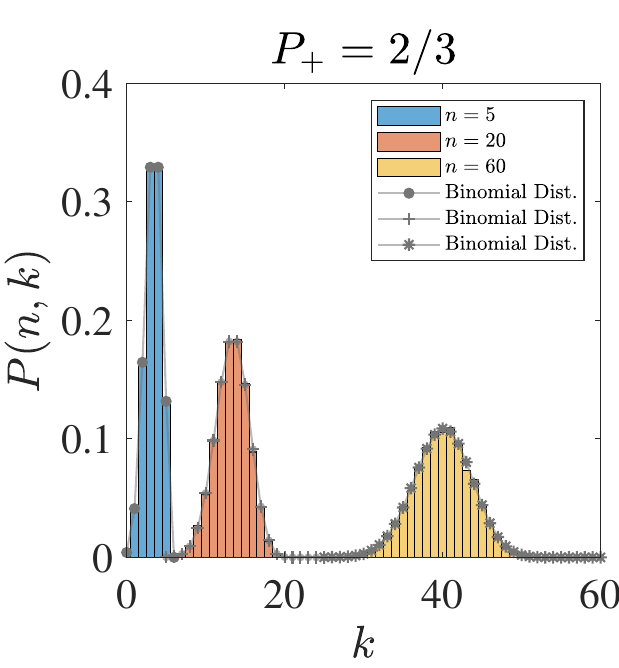}
\put(-118,110){\bf a}~~~
\includegraphics[trim=4.6cm 9.4cm 5.9cm 9.cm, clip=true, scale=0.39, angle=0]{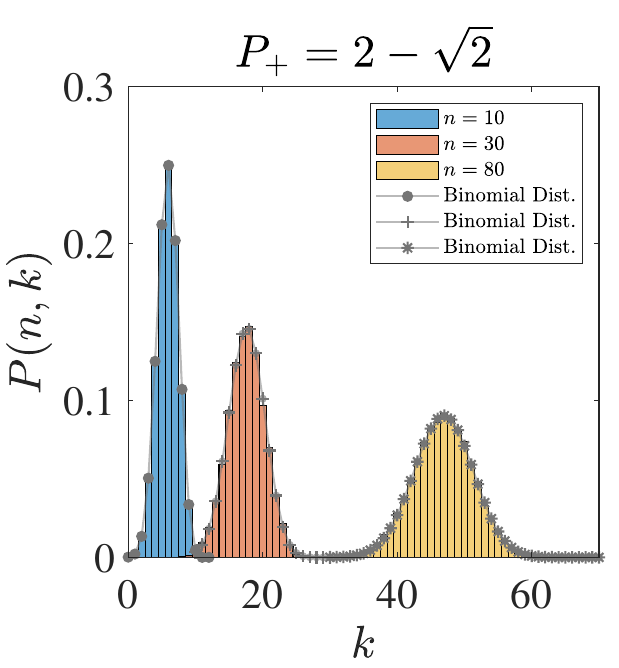}
\put(-118,110){\bf b}\\
\includegraphics[trim=4.6cm 9.4cm 5.9cm 9.cm, clip=true, scale=0.39, angle=0]{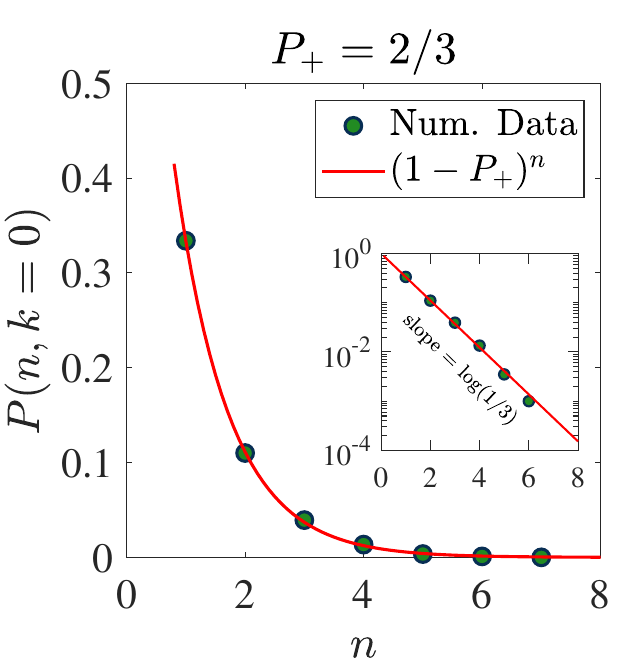}
\put(-118,110){\bf c}~~~
\includegraphics[trim=4.6cm 9.4cm 5.9cm 9.cm, clip=true, scale=0.39, angle=0]{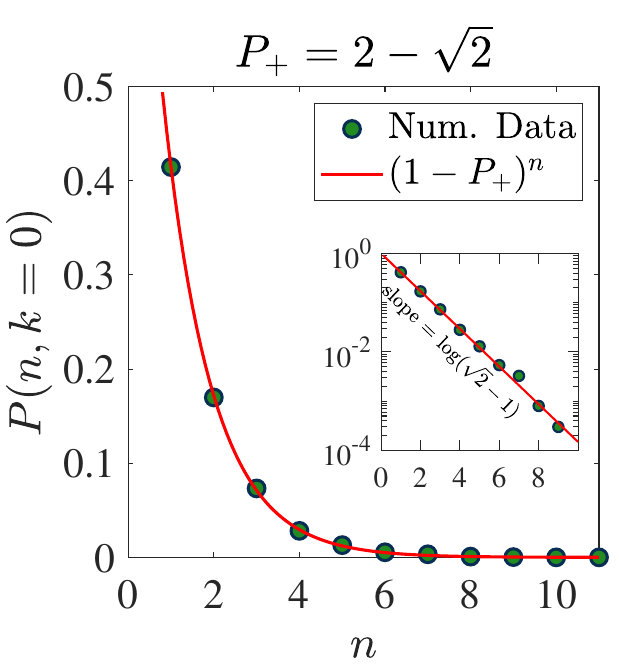}
\put(-118,110){\bf d}
\caption{Counting statistics with Higgs-type potential in Eq.\eqref{higgs}. {\bf a.} Histogram of $P(n, k)$ for three different trial numbers $n$ with the probability $P_+=2/3$; {\bf b.} Histogram of $P(n, k)$ for three different trial numbers $n$ with the probability $P_+=2-\sqrt2$; {\bf c.} The probability of $P(n, k=0)$ in $n$ trials with $P_+=2/3$. The inset is the logarithmic plot, in which the slope of the straight line is $\log(1/3)$, consistent with the prediction $B(n,k=0)=(1-P_+)^{n}$; {\bf d.} The probability of $P(n, k=0)$ in $n$ trials with $P_+=2-\sqrt2$. The inset is the logarithmic plot, in which the slope of the straight line is $\log(\sqrt2-1)$, also matches the prediction $B(n,k=0)=(1-P_+)^{n}$. }\label{histo_GL}
\end{figure}

In Fig.\ref{histo_GL} we show the counting statistics of the rolling down to the right side with the Higgs-type potential. In the left column,  i.e., Fig.\ref{histo_GL} {\bf a} and {\bf c}, we set $\alpha_-=1$ and $\alpha_+=4$. Therefore, according to our theoretical prediction in Eq.\eqref{pzandpf}, the probability of rolling down to the right side is $P_+=\frac{\sqrt{4}}{1+\sqrt{4}}=2/3$. While in the right column, i.e., Fig.\ref{histo_GL} {\bf b} and {\bf d}, we set $\alpha_-=1$ and $\alpha_+=2$, thus $P_+=\frac{\sqrt{2}}{1+\sqrt{2}}=2-\sqrt2$. Panel {\bf a} and {\bf b} of Fig.\ref{histo_GL} show the histogram of the probability $P(n, k)$, which means the probability for $k$ times of rolling down to the right side in $n$ trials. We can see that for different probabilities $P_+$ and different trial numbers $n$, the histogram of $P(n, k)$ are consistent with the theoretical binomial distributions $B(n, k)$ in Eq.\eqref{binodist} very well. We did not show the error bars of the histogram since the errors are very tiny. 

Fig.\ref{histo_GL} {\bf c} and {\bf d} exhibit the extremal cases that there are no events for rolling down to the right side in $n$ trials, i.e., $P(n, k=0)$, for different $P_+$. Theoretically, from the binomial distribution Eq.\eqref{binodist}, this probability for vanishing $k$ should be $B(n, k=0)=(1-P_+)^n$. From Fig.\ref{histo_GL} {\bf c} and {\bf d}, we see that the numerical statistics (circles) match the theoretical predictions (red lines) very well. The inset plots show the relations between $\log P(n,k=0)$ and $n$. From the binomial distribution, the theoretical relation should be $\log B(n,k=0)=n\log(1-P_+)$, which are reflected by the linear relations in the inset plots.  Moreover, the slopes are consistent with the theoretical predictions. Specifically, the slopes are $\log(1-P_+)=\log(1-2/3)=\log(1/3)$ in Fig.\ref{histo_GL} {\bf c} and $\log(1-P_+)=\log(1-(2-\sqrt2))=\log(\sqrt2-1)$ in Fig.\ref{histo_GL} {\bf d}, respectively. 

Therefore, we see that the binomial distributions with the given probability $P_+$ are consistent with the numerical statistics. This in turn implies that our theoretical prediction in Eq.\eqref{pzoverpf} and Eq.\eqref{pzandpf} are correct. 

We must stress that our main findings in Eq.\eqref{pzoverpf} and Eq.\eqref{pzandpf} are robust against by varying other parameters. In numerics, we already checked that they are robust against: (I) varying the reservoir temperature $T$; (II) varying the coefficient $\Gamma$; (III) adding $\phi^3$ terms in the potential Eq.\eqref{higgs}; (IV) changing the coefficients in front of $\phi^4$; (V) adding higher order terms in the potential, such as $\phi^5$ and etc. In the \hyperref[app]{Supplemental Materials } we adopt another cosine-type potential to numerically verifies our theoretical predictions. Therefore, all of these reflect that our predictions in Eq.\eqref{pzoverpf} and Eq.\eqref{pzandpf} are robust and universal.

\section{Conclusions and Discussions}
We studied the unequal probabilities of a field to roll down an asymmetric potential, in which the top is only $C^1$ continuous. By using the Fokker-Planck equation in stochastic process, we theoretically found that the probability to roll down to the left (right) depends on the square root of the second derivative of the potential at that top, i.e. our main conclusions are the Eq.\eqref{pzoverpf} and Eq.\eqref{pzandpf}. Then we used the Higgs-type potentials in the time-dependent Ginzburg-Landau equation to numerically verify our conclusions. We also found that our theoretical predictions are robust against varying other parameters. Therefore, our findings are universal results. 

Our findings are important to the asymmetries of the vacuum selection. Potentially, this new mechanism may play an important role in explaining the origins of asymmetries in the early Universe given that one can build such an asymmetric effective potential first. 


 

\section*{Acknowledgement} The authors thank Peng-Zhang He and Yu Zhou for the helpful discussions. HQZ would like to appreciate many interesting talks, which inspired him to think about this problem, during the conference ``Workshops on Gravitation and Cosmology 2023" held in Beijing by ITP, CAS.  This work was partially supported by the National Natural Science Foundation of China (Grants No.12175008).

\bibliography{refs.bib}

\pagebreak
\widetext

\appendix*

\begin{center}
\section*{---Supplemental Materials---}
\label{app}
\end{center}

\setcounter{equation}{0}
\setcounter{figure}{0}
\setcounter{table}{0}
\setcounter{section}{0}
\makeatletter
\renewcommand{\theequation}{S\arabic{equation}}
\renewcommand{\thefigure}{S\arabic{figure}}
\renewcommand{\bibnumfmt}[1]{[#1]}
\renewcommand{\citenumfont}[1]{#1}

\section{Unequal probabilities of vacuum selection with cosine-type potential}

We will use another potential to verify the correctness of our main findings in Eq.\eqref{pzoverpf} and Eq.\eqref{pzandpf} in the main text. It is the cosine-type potential,  
\be\label{vcos} 
V(\phi)=\theta(-\phi)(\cos(\beta_-\phi)+\phi^4)+\theta(\phi)(\cos(\beta_+\phi)+\phi^4). 
\ee
In numerics, we fix $\beta_-\equiv2$ and vary $\beta_+$ in the range of $\beta_+\in[0.2, 8]$. Now we change the time step to be $\Delta t=0.02$ and still set $\Gamma=5$ and temperature $T=0.01$. 

\begin{figure}[h]
\centering
\includegraphics[trim=4.6cm 9.4cm 4.5cm 9.8cm, clip=true, scale=0.39, angle=0]{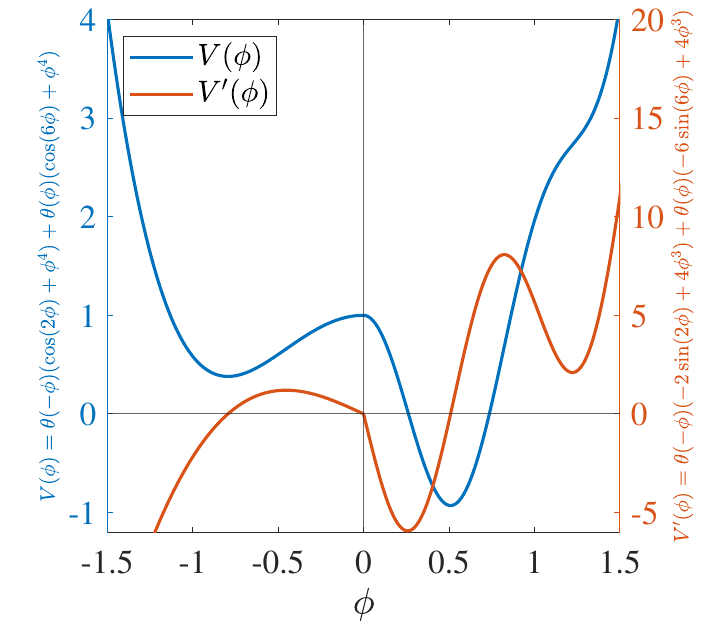}
\put(-128,110){\bf a}~
\includegraphics[trim=4.6cm 9.4cm 5.5cm 9.8cm, clip=true, scale=0.39, angle=0]{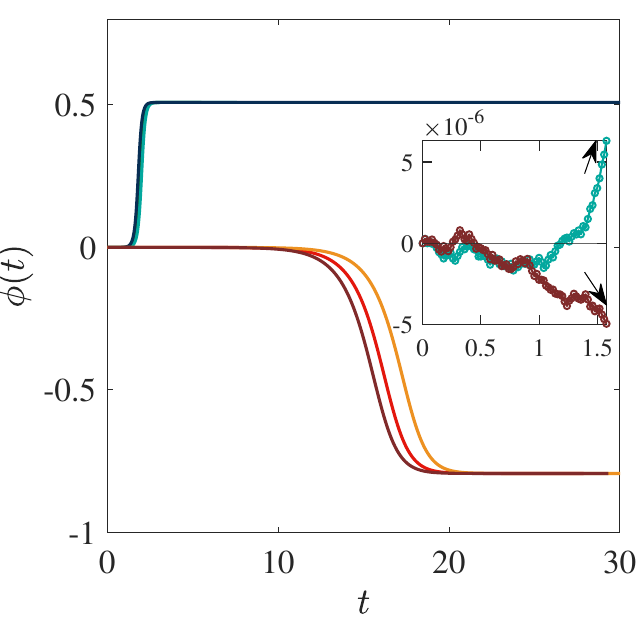}
\put(-118,110){\bf b}
\includegraphics[trim=4.7cm 9.4cm 5.5cm 9.8cm, clip=true, scale=0.39, angle=0]{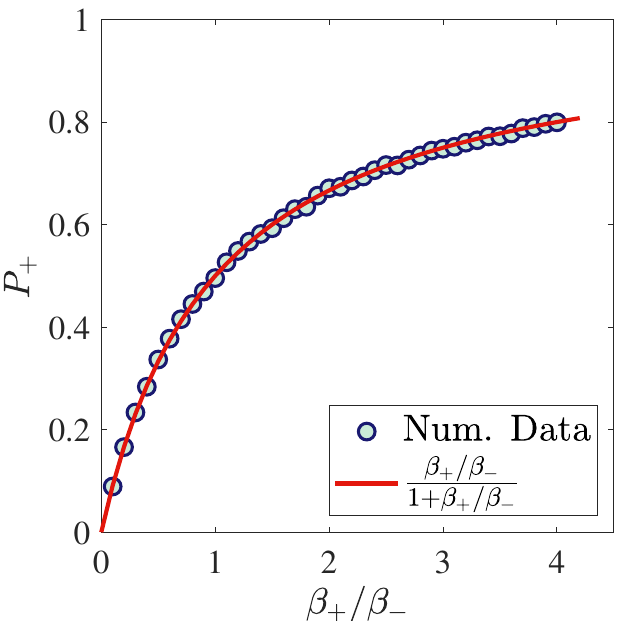}
\put(-118,110){\bf c}
\includegraphics[trim=4.7cm 9.4cm 5.5cm 9.8cm, clip=true, scale=0.39, angle=0]{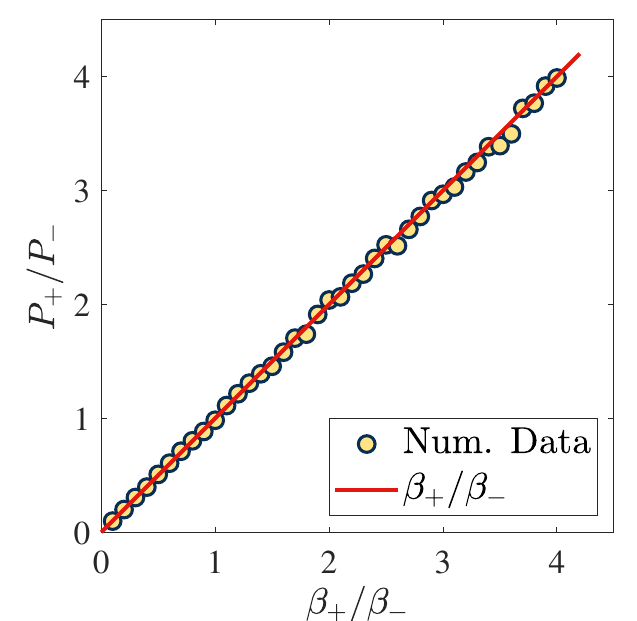}
\put(-118,110){\bf d}
\caption{{\bf a}. The cosine-type potential $V(\phi)$ (blue) in Eq.\eqref{vcos} with $\beta_-=2$ and $\beta_+=6$, and its derivative $V'(\phi)$ (red); {\bf b}. Time time evolution of the scalar fields. Each has three cases to approach positive and negative minimums. The inset plot shows the very early fluctuations around the top due to the temperature fluctuation. After some time, they will persistently roll down the potential as the arrows indicate; {\bf c}. Relation between the probability $P_+$ to the ratio $\beta_+/\beta_-$. Numerical results match the theoretical prediction (red line) Eq.\eqref{pzandpf} in the main text very well; {\bf d}. Relation between the ratio $P_+/P_-$ to the ratio $\beta_+/\beta_-$. The numerical results are completely consistent with the theoretical prediction (red line) in Eq.\eqref{pzoverpf} in the main text. }\label{pzpf_cos}
\end{figure}

Panel {\bf a} of Fig.\ref{pzpf_cos} shows the profiles and the derivative of the potential \eqref{vcos} with $\beta_+=6$. Obviously, the potential satisfies our requirement of the $\mathring{C}^1$ continuous. The minimum values of potential sits at $\phi_-=-0.7937$ and $\phi_+=0.5089$. 

Panel {\bf b} of Fig.\ref{pzpf_cos} exhibits the time evolutions of the scalar fields.  There are respectively three cases for each of them to approach $\phi_+$ (the upper lines) and $\phi_-$ (the lower lines) at equilibrium time.  The inset plot shows the initial fluctuations of the scalar fields around the top, and then continuously roll down the potential as the arrows indicate.

Panel {\bf c} of Fig.\ref{pzpf_cos} is the relation between $P_+$ and the ratio $\beta_+/\beta_-$. From the potential \eqref{vcos} we see that $\beta_+/\beta_-=\sqrt{V''_+/V''_-}$, therefore, the relations shown in panel {\bf c}, i.e., $P_+=(\beta_+/\beta_-)/(1+\beta_+/\beta_-)$, exactly satisfies our prediction in Eq.\eqref{pzandpf} in the main text. 

Panel {\bf d} of Fig.\ref{pzpf_cos} shows the relations between $P_+/P_-$ and $\beta_+/\beta_-$. The linear relation $P_+/P_-=\beta_+/\beta_-$ in numerics and theoretical predictions Eq.\eqref{pzoverpf} in the main text match each other very well. 

\section{Binomial distribution with cosine-type potential}
\begin{figure}[h]
\centering
\includegraphics[trim=4.6cm 9.4cm 5.9cm 9.cm, clip=true, scale=0.39, angle=0]{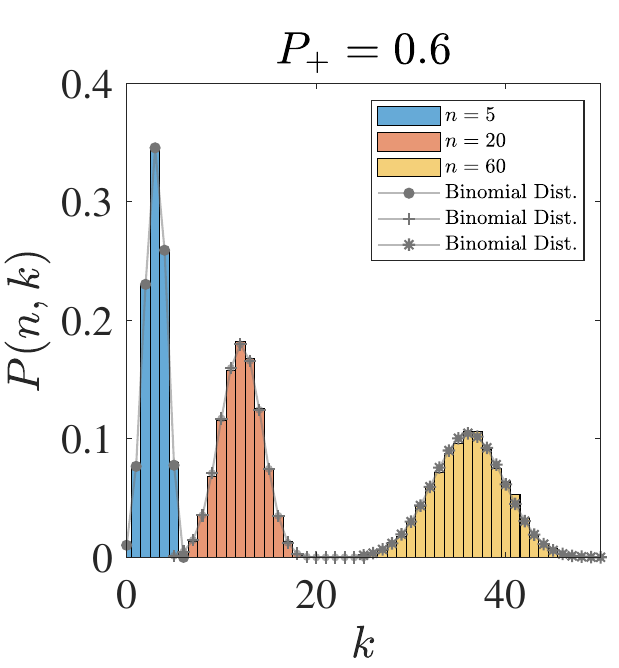}
\put(-118,110){\bf a}~
\includegraphics[trim=4.6cm 9.4cm 5.9cm 9.cm, clip=true, scale=0.39, angle=0]{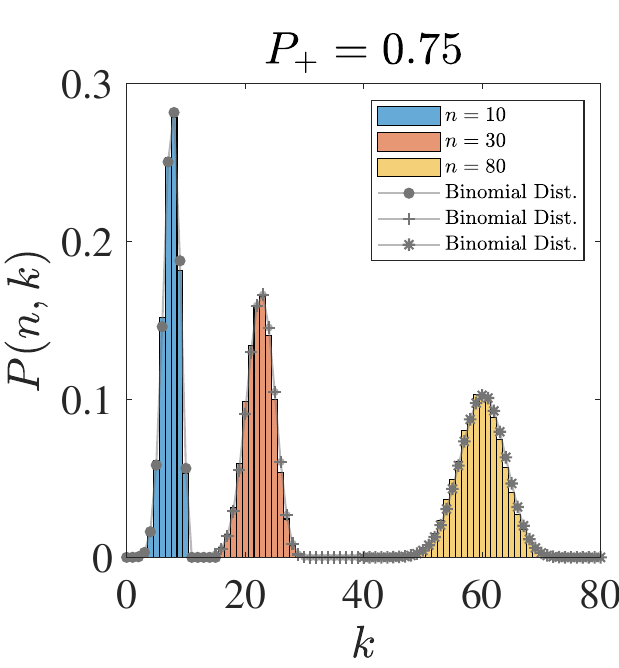}
\put(-118,110){\bf b}~
\includegraphics[trim=4.6cm 9.4cm 5.9cm 9.cm, clip=true, scale=0.39, angle=0]{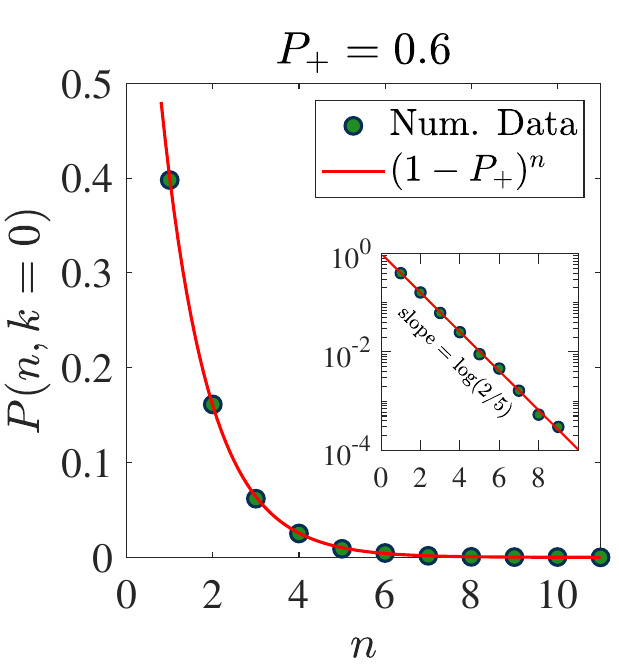}
\put(-118,110){\bf c}~
\includegraphics[trim=4.6cm 9.4cm 5.9cm 9.cm, clip=true, scale=0.39, angle=0]{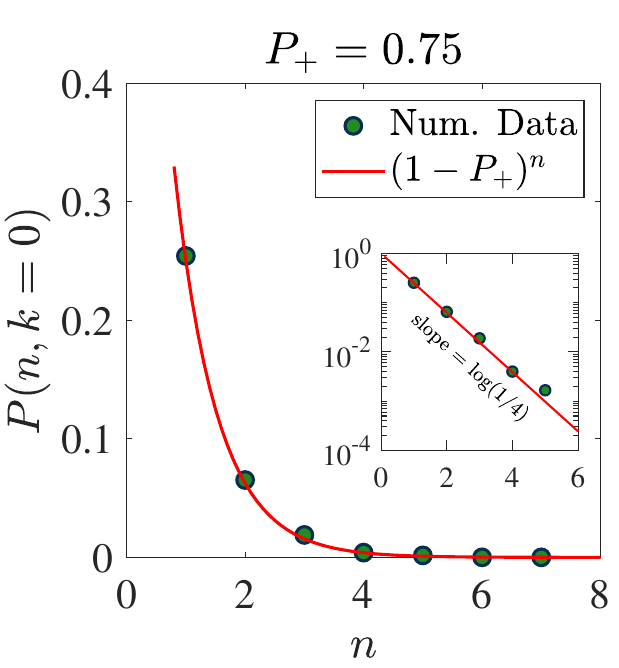}
\put(-118,110){\bf d}
\caption{Counting statistics with cosine-type potential in Eq.\eqref{vcos}. {\bf a.} The histogram of $P(n, k)$ for three different trial numbers $n$, with the probability $P_+=3/5$; {\bf b.} The histogram of $P(n, k)$ for three different trial numbers $n$, with the probability $P_+=3/4$; {\bf c.} The vanishing probability of $P(n, k=0)$ in $n$ trials with $P_+=3/5$. The inset is the logarithmic plot, in which the slope of the straight line is $\log(2/5)$, consistent with the prediction $B(n,k=0)=(1-P_+)^{n}$; {\bf d.} The vanishing probability of $P(n, k=0)$ in $n$ trials with $P_+=3/4$. The inset is the logarithmic plot, in which the slope of the straight line is $\log(1/4)$, consistent with the prediction $B(n,k=0)=(1-P_+)^{n}$.
}\label{histo_cos}
\end{figure}

In Fig.\ref{histo_cos} we show the counting statistics of the rolling down to the right side with the cosine-type potential in Eq.\eqref{vcos}. For the left column,  i.e., Fig.\ref{histo_cos} {\bf a} and {\bf c}, we set $\beta_-=2$ and $\beta_+=3$. Therefore, the probability to roll down to the right side is $P_+=\frac{3/2}{1+3/2}=3/5$ according to Eq.\eqref{pzandpf} in the main text. While for the right column, i.e., Fig.\ref{histo_cos} {\bf b} and {\bf d}, we set $\beta_-=2$ and $\beta_+=6$, thus $P_+=\frac{6/2}{1+6/2}=3/4$. 

From panel {\bf a} and {\bf b} of Fig.\ref{histo_cos}, we can see that for the different probabilities $P_+$ and different trials numbers $n$, the histogram of $P(n, k)$ are consistent with the theoretical binomial distributions $B(n, k)$ in Eq.\eqref{binodist} in the main text very well. 

Fig.\ref{histo_cos} {\bf c} and {\bf d} shows the probability to roll down to the right with zero times in $n$ trials, i.e., $P(n, k=0)$. Theoretically, from the binomial distribution Eq.\eqref{binodist} in the main text, this probability should be $B(n, k=0)=(1-P_+)^n$. The two inset plots show the corresponding logarithmic figures, i.e., $\log P(n,k=0)$ vs. $n$.   Specifically, from the inset plots we can see that the slopes are exactly the value of $\log(1-P_+)$ since $\log B(n,k=0)=n \log(1-P_+)$. Therefore, from the two plots we see that the numerical statistics match the theoretical predictions very well. 

Consequently, we find that by using the cosine-type potential \eqref{vcos}, the consistency between the numerics and the binomial distributions in turn implies that our main predictions in Eq.\eqref{pzoverpf} and Eq.\eqref{pzandpf} in the main text are correct.

\end{document}